Highly resistive epitaxial Mg-doped GdN thin films

C.-M. Lee,[1] H. Warring,[1] S. Vézian,[2] B. Damilano,[2] S. Granville,[3] M. Al Khalfioui,[2,4] Y. Cordier,[2] H..J. Trodahl,[1] B.J. Ruck,[1] F. Natali,[1,a)]

[1] MacDiarmid Institute for Advanced Materials and Nanotechnology, School of Chemical and Physical Sciences, Victoria University of Wellington, PO Box 600, Wellington, New Zealand
[2] Centre de Recherche sur l'Hétéro-Épitaxie et ses Applications (CRHEA), Centre National de la Recherche Scientifique, Rue Bernard Gregory, 06560 Valbonne, France
[3] MacDiarmid Institute for Advanced Materials and Nanotechnology, Robinson Research Institute, Victoria University of Wellington, PO Box 33436, Lower Hutt 5046, New Zealand
[4] University of Nice Sophia Antipolis, Parc Valrose, 06102 Nice Cedex 2, France

We report the growth by molecular beam epitaxy of highly resistive GdN, using intentional doping with magnesium. Mg-doped GdN layers with resistivities of $10^3$ $\Omega$.cm and carrier concentrations of $10^{16}$ cm$^{-3}$ are obtained for films with Mg concentrations up to $5 \times 10^{19}$ atoms/cm$^3$. X-ray diffraction rocking curves indicate that Mg-doped GdN films have crystalline quality very similar to undoped GdN films, showing that the Mg doping did not affect the structural properties of the films. A decrease of the Curie temperature with decreasing the electron density is observed, supporting a recently suggested magnetic polaron scenario [F. Natali *et al.*, Phys. Rev. B **87**, 035202 (2013)].

---

a) Electronic mail: franck.natali@vuw.ac.nz

The rare-earth mononitrides (REN) have been recognised for decades to be ferromagnetic at low temperature.[1,2] However, their propensity for oxidation in air, combined with a tendency for the formation of nitrogen vacancies, ensured that their band structure and electron transport were poorly characterised. Recent advances in the growth and passivation of epitaxial films have begun to clarify these properties, and most now appear to be semiconductors.[2] They thus contribute new members to the relatively sparsely populated class of *intrinsic* ferromagnetic semiconductors, lending them both fundamental and technological importance with already proof of concept REN-based device structures.[3,4,5,6,7] The most thoroughly studied REN is GdN, interesting for its purely spin magnetic moment in the half-filled 4$f$ shell of $Gd^{3+}$. It has also the highest Curie temperature ($T_C$) of the series, reported as 70 K in most studies of the past 50 years. However a recent report suggests that $T_C$ has been enhanced by magnetic polarons from less than 50 K in undoped films.[8] Nominally undoped epitaxial GdN thin films typically show an n-type conductivity, primarily due to doping by nitrogen vacancies ($V_N$).[2] A literature study shows that there is a wide range of epitaxial GdN thin films reported with electron carrier densities typically above $10^{20}$ cm$^{-3}$.[8,9,10,11] So far, despite thin-film growth advances, it has remained difficult to reduce the concentration of $V_N$ much below the 1% level, which dopes the films with as many as three electrons per $V_N$. Such relatively high $V_N$ concentrations will remain a problem, in view of the epitaxial growth temperature, typically above 500°C, and the small $V_N$ formation energy.[2] From a device perspective it will be crucial to decrease the conductivity. We have thus explored the compensation of the electron doping of GdN from the nitrogen vacancies by the introduction of $Mg^{2+}$ acceptors.

The GdN:Mg layers were grown by molecular beam epitaxy (MBE) on 100 nm thick wurtzite (0001) oriented high resistive AlN buffer layers which were themselves grown by MBE on (111) silicon substrates. To prevent decomposition in air the GdN:Mg layers were capped with a 100-150 nm thick GaN layer. Atomic nitrogen species were produced by the thermally activated decomposition of ammonia ($NH_3$) on the growing surface while conventional effusion cells were used for Al, Ga, Mg and Gd solid sources. The GdN:Mg layers with thicknesses ranging from 100 nm to 140 nm were grown at a substrate temperature of ~650°C and the $NH_3$ and Gd beam equivalent pressures (BEP) were $1.9\times10^{-5}$ Torr and $5\times10^{-8}$ Torr, respectively. The BEP of magnesium ranged from $5\times10^{-11}$ to $6.5\times10^{-8}$ Torr corresponding to temperatures of the Mg effusion cell between 162°C and 300°C. The GdN:Mg growth rate was ~0.15μm/h. A typical cross-sectional scanning electron microscope image of the structure grown in this letter is reported in Figure1(a). The Mg concentration was determined by secondary ion mass spectrometry (SIMS). The crystalline structure was assessed by X-ray diffraction (XRD) 2θ scans and rocking curves. Electrical transport studies were carried out, including Hall effect measurements at room temperature and temperature dependent resistivity measurements performed in a closed-cycle cryostat between room temperature and 4 K. The magnetic properties of the films were investigated using a superconducting quantum interference device (SQUID) magnetometer.

The Mg concentration depth profile in the GdN layers, as measured by SIMS, is displayed in Figure 1(b). For the two Mg BEPs $1.3\times10^{-9}$ and $8.0\times10^{-9}$ Torr, the Mg concentration was $1\times10^{19}$ and $5\times10^{19}$ atoms/cm$^3$, changing as expected linearly with the incident Mg flux, or exponentially with the inverse of the effusion cell temperature. The Mg concentration has a rather flat profile, suggesting the dopant incorporation rate was constant during the growth. The upturn at the film surface is believed to be caused by incorporation of residual Mg in the chamber prior to deposition of the capping layer. The crystalline order/quality of GdN layers doped up to $5\times10^{19}$ atoms/cm$^3$ is very similar to that of an undoped GdN layer grown under the same conditions as shown in Figure 1(c). The (111) XRD rocking curve full width at half maximum (FWHM) is 7200 arsec for both a 140 nm thick Mg-doped GdN layer with a concentration of $5\times10^{19}$ Mg atoms/cm$^3$ and an undoped GdN layer. The FWHM of the GdN (111) reflections from 2θ/ω scans are very similar for undoped samples and samples doped up to $5\times10^{19}$ Mg atoms/cm$^3$. This shows that the Mg doping did not significantly affect the structural properties of the film. However, the quality of the material does start to deteriorate for a Mg BEP of $6.5\times10^{-8}$ Torr, corresponding to a Mg concentration exceeding $2\times10^{20}$ Mg atoms/cm$^3$ as obtainedby extrapolating the SIMS results.

Encouraged by the promising growth results, resistivity and Hall effect measurements were performed at room temperature. The efficacy of compensation by Mg can be immediately seen in the ambient-temperature electron carrier concentrations over the full range of Mg cell temperatures. The electron density of $7 \times 10^{20}$ cm$^3$ for an undoped GdN layer can thus be reduced by as much as five orders of magnitude down to $\sim 5 \times 10^{15}$ cm$^3$ for a doped GdN layer [Figure 2(a)]. For Mg concentrations above $5\times10^{19}$ Mg atoms/cm$^3$ the GdN:Mg films are highly resistive (>1000 Ω.cm). Figure 2(b) shows that the room temperature resistivity varies inversely with the electron density over five orders of magnitude, implying an approximately constant electron mobility of $\sim 5$ cm$^2$/Vs over the full range of Mg concentration.

We focus now on two films with Mg concentrations of 0 (sample *H*) and $10^{19}$ atoms/cm$^3$ (sample *M*) having ambient-temperature Hall carrier concentrations of ~7 x $10^{20}$ and ~5 x $10^{16}$ cm$^{-3}$ and ambient-temperature resistivities of 0.0024 and 25 $\Omega$.cm, respectively. Note that the concentration in the undoped film corresponds to 0.02 electrons per primitive cell, suggesting that the relatively low mobilities are due to a $V_N$ concentration of ~1%.

It appears at first glance improbable that a Mg concentration of $10^{19}$ atoms/cm$^3$ should passivate so fully an electron concentration of nearly $10^{21}$ cm$^{-3}$. It is clearly not a matter of simply compensating the $V_N$ donors, rather it is likely that the network is altered sufficiently that there is a reduced $V_N$ concentration when the films are grown with Mg as a dopant. Note, however, that there is recent evidence that the electrons are found in shallow traps near the $V_N$, forming magnetic polarons for $V_N$ concentrations above ~$10^{20}$ cm$^{-3}$.[8] The magnetic data described below strongly supports this model.

The temperature dependence of the resistivities of the two films are shown in Fig. 3. The resistivity of the undoped film shows a positive temperature coefficient of resistance (TCR) near room temperature, as is commonly seen in heavily doped epitaxial GdN films. Clearly the 0.02 electrons per primitive cell form a degenerate electron gas, though it is as yet unclear whether they reside in the conduction band (CB) or in a defect-centred tail below the CB edge. There is an anomaly peaking very close to the 70 K Curie temperature in this film (see magnetic measurements below) which is likely associated with magnetic disorder scattering.[8] In contrast the Mg doped film has not only a semiconducting magnitude of the resistance, but also shows a negative TCR typical of a semiconductor but with a small activation energy. In this case the Curie temperature ($T_C$) is close to 55 K (see below) and the anomaly is again found near $T_C$. This sample shows also a negative TCR at the lowest temperatures indicating that the ferromagnetic phase is also semiconducting.

The field-cooled (FC) magnetisation measurements in Figure 4 show that both samples are ferromagnetic at low temperature, but with substantial contrasts. The undoped film (sample *H*) (black squares) appears to transform homogeneously in the ferromagnetic phase at ~70 K; indeed even its paramagnetic response plotted in the insert demonstrates that homogeneous transformation, for all of the 7 µB Gd ions participate in the 70 K TC Curie-Weiss fit. The Mg-doped film (sample *M*) also shows a very weak onset of ferromagnetism at 70 K, but in this case the majority of the $Gd^{3+}$ ions order only at the lower Curie temperature of ~45 K**.** An analogous FC data set for a third film, with even higher Mg concentration (5 x $10^{19}$ $cm^{-3}$, i.e. a carrier concentration of ~6 x $10^{15}$ $cm^{-3}$) shows a further reduction in the magnetic response in the foot between 45 and 70 K, exactly as expected in the magnetic polaron scenario.[8,12,13]

In conclusion, Mg-doped GdN epilayers were grown by molecular beam epitaxy. Electrical measurements clearly show compensation of the high electron density of GdN by the introduction of magnesium atoms. No deterioration of the structural properties is observed for Mg concentration up to $5 \times 10^{19}$ $atoms/cm^3$. A significant reduction of the electron density down to $10^{16}$ $cm^{-3}$ and increase of the resistivity up to $10^3$ Ω.cm is observed for a Mg concentration of $5 \times 10^{19}$ $atoms/cm^3$. A decrease of the Curie temperature when decreasing the electron density is observed supporting a magnetic polaron scenario.

We acknowledge funding from the Marsden Fund (Grant No. 13-VUW-1309), and the MacDiarmid Institute for Advanced Materials and Nanotechnology, funded by the New Zealand Centres of Research Excellence Fund. We acknowledge support from GANEX (ANR-11-LABX-0014). GANEX belongs to the public funded "Investissements d'Avenir" program managed by the French ANR agency.

## Figures and Figure Captions

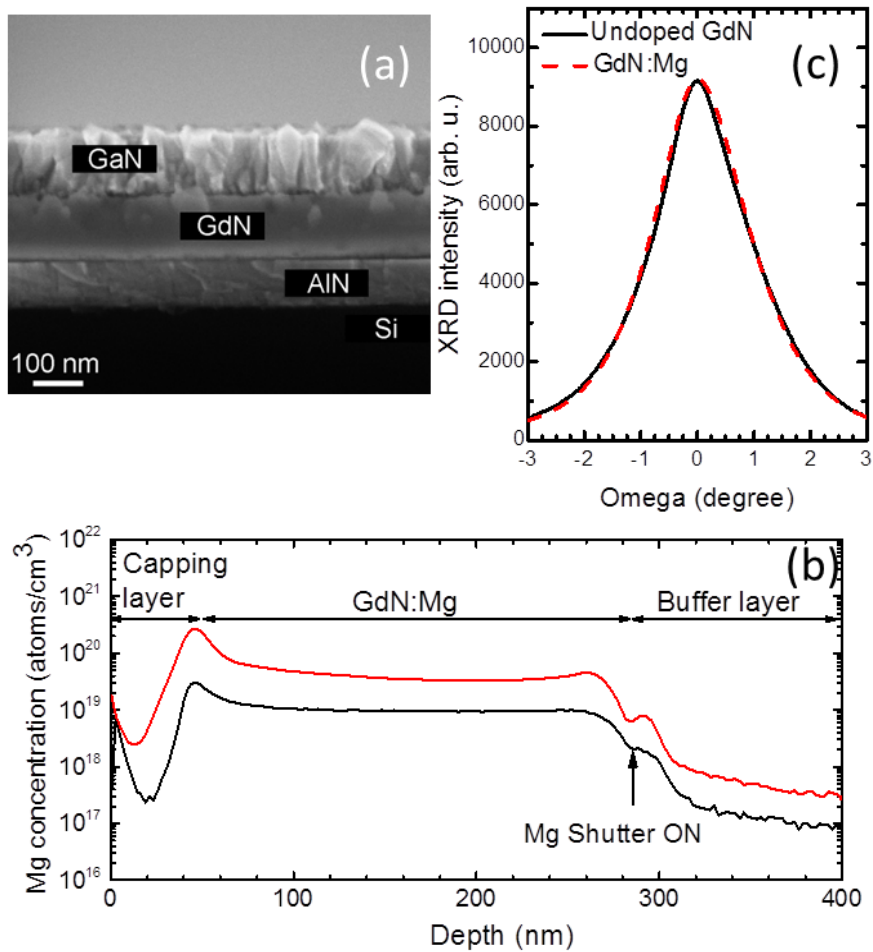

Figure 1: (a) Cross sectional SEM image showing the grown structure. (b) SIMS profiles of Mg in GdN:Mg layers grown with Mg BEPs of $1.3\times10^{-9}$ (black) and $8.0\times10^{-9}$ Torr (red). The measured Mg concentrations are about $1\times10^{19}$ and $5\times10^{19}$ atoms/cm$^3$. (c) Rocking curves of an 140 nm thick undoped GdN layer (black line) and doped GdN layer (red dashed line) with a concentration of $5\times10^{19}$ Mg atoms/cm$^3$.

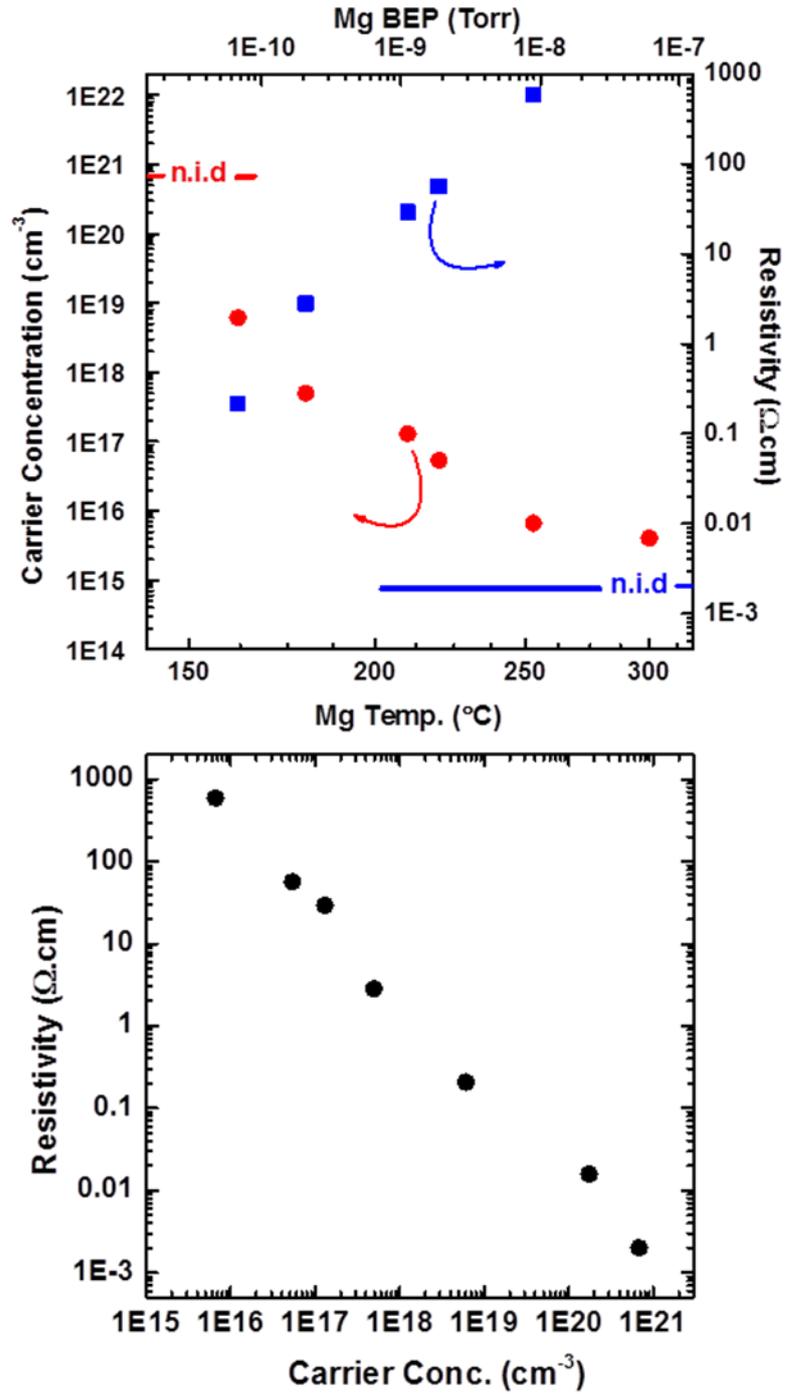

Figure 2: (a) Electron concentration (circles) and resistivity (squares) of 100 nm thick GdN:Mg layers as a function of the Mg effusion cell temperature. The resistivity and carrier concentration for undoped GdN layers are by the n.i.d lines (b) Resistivity of 100 nm thick GdN:Mg layers as a function of the electron concentration.

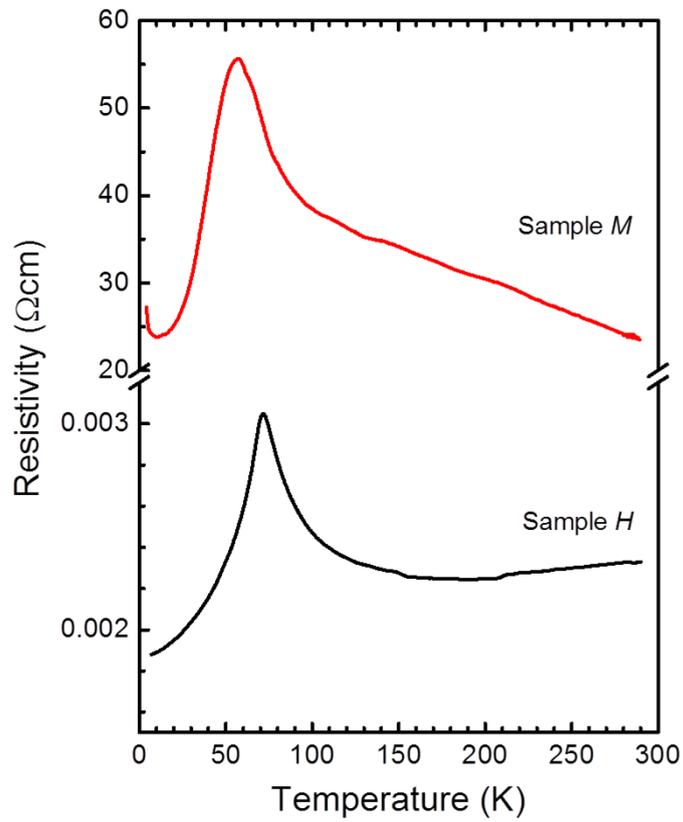

Figure 3: Temperature dependent resistivity of an undoped GdN film (sample *H*) and a GdN:Mg layer doped with a Mg concentration of $10^{19}$ atoms/cm$^3$ (sample *M*)

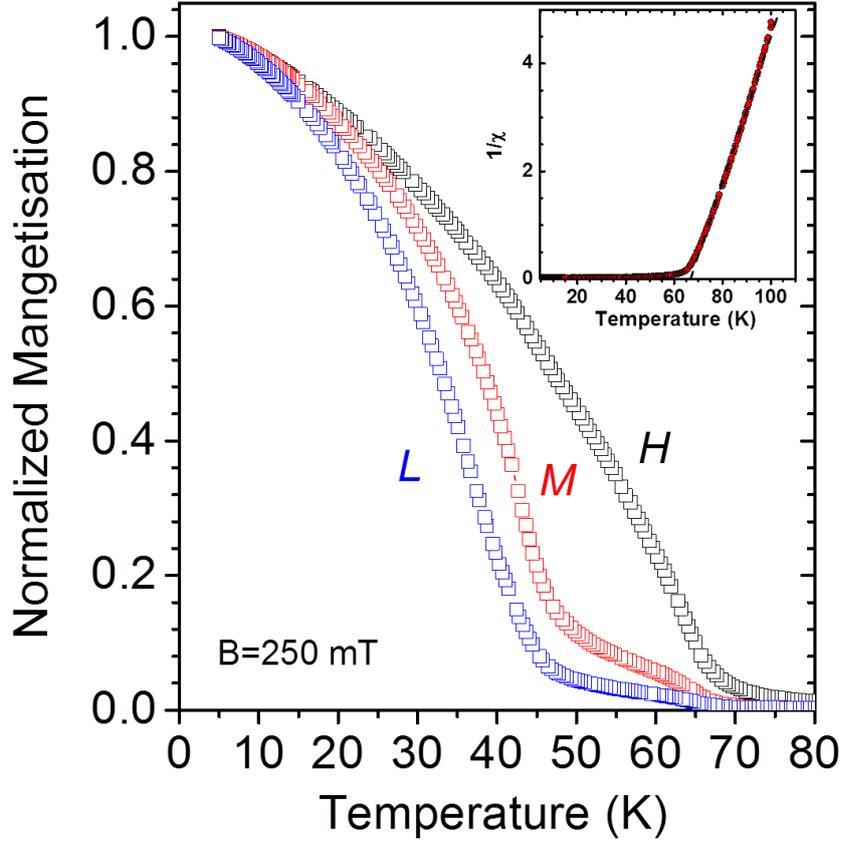

Figure 4: Temperature dependent magnetisation of GdN films with electron carrier concentrations of ~ 6 x $10^{15}$ cm$^{-3}$ (sample *L*), ~5 x $10^{16}$ cm$^{-3}$ (sample *M*), and ~7 x $10^{20}$ cm$^{-3}$ (sample *H*). Sample *H* is an undoped sample, samples *M* and *L* correspond to GdN:Mg layers doped with a Mg concentration of $1\times10^{19}$ and $5\times10^{19}$ atoms/cm$^3$, respectively. The inset is the temperature-dependent inverse susceptibility in an applied field of 250 mT for the undoped film (sample *H*).